\documentclass[aps, pra, twocolumn, nofootinbib, superscriptaddress]{revtex4-2}

\usepackage{comment}
\usepackage{color}
\usepackage{cancel}
\usepackage{ulem}

\usepackage{graphicx}

\usepackage{amsfonts}
\usepackage{amsmath}

\usepackage{physics}
\usepackage{mathrsfs}  
\newtheorem{theorem}{Theorem}

\begin{document}


\title{Selection Rule for Topological Amplifiers in Bogoliubov-de Gennes Systems} 
\author{Hong Y. Ling}
\affiliation{Department of Physics and Astronomy, Rowan University, Glassboro, New Jersey 08028, USA}
\author{Ben Kain}
\affiliation{Department of Physics, College of the Holy Cross, Worcester, Massachussets 01610, USA}

\date{\today}

\begin{abstract}

 Dynamical instability is an inherent feature of bosonic systems described by the Bogoliubov-de Gennes (BdG) Hamiltonian. Since it causes the BdG system to collapse, it is generally thought that it should be avoided.  Recently, there has been much effort to harness this instability for the benefit of creating a topological amplifier with stable bulk bands but unstable edge modes which can be populated at an exponentially fast rate.  We present a theorem for determining the stability of states with energies sufficiently away from zero, in terms of an unconventional commutator between the number conserving part and number nonconserving part of the BdG Hamiltonian. We apply the theorem to a generalization of a model from Galilo et al. [Phys. Rev. Lett, 115, 245302(2015)] for creating  a topological amplifier in an interacting spin-1 atom system in a honeycomb lattice through a quench process.  We use this model to illustrate how the vanishing of the unconventional commutator selects the symmetries for a system so that its bulk states are stable against (weak) pairing interactions.  We find that as long as  time reversal symmetry is preserved, our system can act like a topological amplifier, even in the presence of an onsite staggered potential which breaks the inversion symmetry.
 
\end{abstract}

\pacs{Valid PACS appear here}
\maketitle


\section{Introduction}
The hallmark of topological matter is the presence of gapless  edge modes that support the flow of currents along its boundaries in a manner immune to impurity scattering \cite{hasan10RevModPhys.82.3045,qi2011RevModPhys.83.1057}. This property, which is as novel and fascinating as current flow without resistance in superconductivity \cite{bardeen57PhysRev.108.1175}, follows from the nontrivial topology of Bloch band structures.  As the Bloch band is a universal feature of any periodic systems, topological phenomena are ubiquitous, attracting intense interest across broad areas of research,  including ultracold atoms \cite{aidelsburger2013NatPhys.9.795,aidelsburger2013PhysRevLett.111.185301,hirokazu2013PhysRevLett.111.185302,jotzu2014Nature.515.237,Stuhl2015Science.349.1514}, photons \cite{haldane08PhysRevLett.100.013904,wang2009Nature.461.772,hafezi2011NaurePhysics.7.907,kejie2012NatPhoton2012}, phonons (and mechanical metamaterials) \cite{kane2013NaturePhysics.10.39,prodan2009PhysRevLett.103.248101,fleury2014Science.343.516,yang2015PhysRevLett.114.114301,peano2015PhysRevX.5.031011,he2016NatPhys.12.1124}, and magnons \cite{shindou2013PhysRevB.87.174402,shindou2013PhysRevB.87.174427}
 (see recent review articles \cite{cooper2019RevModPhys.91.015005,ozawa2019RevModPhys.91.015006,zhang2018CommunPhys.1.97,kondo2020} and references therein).  This ubiquity also means that systems need no longer be closed, fermionic,  Hermitian, and in thermal equilibrium in order to exhibit topological phases.  Indeed,  significant effort in recent years has been devoted to investigating open systems described by non-Hermitian Hamiltonians \cite{esaki2011PhysRevB.84.205128,liang2013PhysRevA.87.012118,lee2016PhysRevLett.116.133903,leykam2016PhysRevLett.117.143901,xiong2018JPhysCommun.2.035043,shen2018PhysRevLett.120.146402,yao2018PhysRevLett.121.086803} (see \cite{gong2018PhysRevX.8.031079,kawabata2019PhysRevX.9.041015,zhou2019PhysRevB.99.235112,ghatak2019JOfPhysCondMatt.31.263001,ashida2020} and references therein).

In this paper, we consider quadratic bosonic systems \cite{barnett13PhysRevA.88.063631,shindou2013PhysRevB.87.174427,galilo15PhysRevLett.115.245302,engelhardt2015PhysRevA.91.053621,engelhardt2016PhysRevLett.117.045302,bardyn2016PhysRevB.93.020502,peano2016PhysRevX.6.041026,furukawa2017NewJournalOfPhysics.17.115014,shindou2013PhysRevB.87.174427}.  A new possibility arises in systems with bosons: Unrestricted by the Pauli exclusion principle, bosonic systems may be made to operate as active topological matter (a topological ``laser") where population amplification occurs only in modes localized along edges. Schomerus \cite{schomerus2013OptLett.38.1912} sought to achieve this novel phase in a photonic crystal realization \cite{poli2015NatCommun.6.6710} of the Su-Schrieffer-Heeger model \cite{su1979PhysRevLett.42.1698}, which is a non-Hermitian system with gain and loss.  We focus on an alternative possibility \cite{barnett13PhysRevA.88.063631,galilo15PhysRevLett.115.245302,engelhardt2016PhysRevLett.117.045302,peano2016PhysRevX.6.041026} which seeks to achieve the same goal in systems described by a quadratic bosonic Hamiltonian with pairing terms, which is the bosonic analog of the Bogoliubov-de Gennes (BdG) Hamiltonian for fermions.  

 The bosonic BdG system is Hermitian, but, unlike the fermionic BdG model, the matrix to be diagonalized is non-Hermitian. Hence, the bosonic BdG model inherits a so-called dynamical instability, which occurs when any subset of eigenvalues become complex.  
Recent proposals for realizing a topological amplifier in light \cite{peano2016PhysRevX.6.041026} and in matter \cite{barnett13PhysRevA.88.063631,galilo15PhysRevLett.115.245302,engelhardt2016PhysRevLett.117.045302} make use of this dynamical instability. 

This approach is not uncommon, as many exciting phenomena in physics are built on instabilities. For example, the production of
highly coherent light in a laser follows from an instability
associated with population inversion between upper and
lower energy levels \cite{lamb1974laserPhysics} and the simultaneous amplification
of signal and idler modes in light \cite{milburn1981OptCommun39.401,slusher1985PhysRevLett.55.2409} and in matter \cite{leslie2009PhysRevA.79.043631,campbell2006PhysRevLett.96.020406,klempt2010PhysRevLett.104.195303}
 follows from a parametric instability \cite{mandel1995} which
occurs when the strength of the driving field exceeds a
critical threshold. 

 A particular challenge with creating a topological amplifier lies in the dilemma that, on one hand, the topological amplifier must have an instability-free bulk (or, at least, a bulk far more stable than its edges) but, on the other hand, the dynamical instability in the bosonic BdG system tends to be impartial to edge and bulk states.  An immediate question is how to control the instability so that it affects edge modes but not bulk modes.  Answering this question is central to a successful implementation of topological amplifiers in BdG systems. The goal of the present work is to formulate a systematic approach to this question.

Our paper is organized as follows.  In Sec. \ref{sec:a theorem}, we consider a bosonic system described by a prototypical BdG Hamiltonian made up of number conserving (``normal") and number nonconserving (``anomalous," i.e., pairing) parts. It is well-known \cite{wu2001PhysRevA.64.061603,kawaguchi2004PhysRevA.70.043610,nakamura2008PhysRevA.77.043601}
that complex eigenvalues and hence the dynamical instability in a bosonic BdG system can be traced to level crossings between particle and hole states in the absence of the pairing interaction.  We thus focus on a pair of degenerate particle and hole states and investigate how the pairing interaction, which is treated as a perturbation, lifts the degeneracy.  We combine first-order degenerate perturbation theory with the bi-orthonormality condition to establish a theorem for determining the stability of a state with energy sufficiently far from zero in a BdG system.  Recognizing that the heart of the matter is level crossings between particle and hole states, we express this theorem in terms of an unconventional commutator between number conserving and number nonconserving parts of a BdG Hamiltonian [see Eq. (\ref{eq:a commutator})]. We show that when this ``commutator" vanishes, the first-order splitting, which is purely imaginary, disappears and the two states remain stable and degenerate. We further show that when this  ``commutator" vanishes, the energy corrections remain real not only at second order in Sec.  \ref{sec:a theorem} but also to all higher orders in Appendix \ref{sec:Higher than second-order energy corrections}.  We thus identify the vanishing of the unconventional commutator (\ref{eq:a commutator}) as a general and straightforward to use guiding principle for creating a topological amplifier in BdG systems.

 In Appendix \ref{appendix 1}, we describe a generalization of a model from Galilio et al. \cite{galilo15PhysRevLett.115.245302} for creating a topological amplifier in an interacting spin-1 atom system in a honeycomb lattice through a quench process. In Sec. \ref{sec:application}, we apply our theorem to this generalized model which is neither time reversal nor inversion symmetric and we use it to demonstrate how the principle we developed in Sec.\ref{sec:a theorem} selects the symmetries that a BdG system must possess so that it may behave like a topological amplifier. We find that as long as time reversal symmetry is preserved, our system can act like a topological amplifier, even in the presence of an onsite staggered potential which breaks the inversion symmetry. We conclude in Sec. \ref{sec:conclusion}.
 
\section{A stability theorem for bosonic B\lowercase{d}G systems}
\label{sec:a theorem}

A topological amplifier in the context of the present work must have stable bulk states and unstable edge states. Then, a first step towards engineering such an amplifier is a good understanding of how a state in a BdG system can become unstable. In this section, we present a theorem which allows us to determine the stability of a state with energy sufficiently far from zero in a BdG system.  Let such a (finite) system be described by the BdG Hamiltonian in Nambu space,
\begin{equation}
\label{eq:H_BdG}
    H_\mathrm{BdG}(k) = 
    \begin{pmatrix}
     A(k) & B(k)\\
     -B^*(-k) & -A^*(-k)
    \end{pmatrix},
\end{equation}
where $k$ stands for the crystal momentum (which may contain several components) within the first Brillouin zone.  Equation (\ref{eq:H_BdG}) is not necessarily Hermitian,  but since it is the BdG Hamiltonian for a Hermitian system, $\Sigma_zH_\mathrm{BdG}(k)$ must be Hermitian, where the matrix $\Sigma_z$ is defined directly below Eq. (\ref{eq:particle-hole symmetry}). Consequently,  
\begin{align}
\label{eq:A and B Properties}
A(k) = & A^\dag(k), \qquad B(k) =B^T(-k),
\end{align}
 which are square matrices in an internal space and describe, respectively, number conserving and nonconserving (or pairing) processes. The BdG Hamiltonian (\ref{eq:H_BdG}) has two properties: so-called pseudo-Hermiticity,
\begin{equation}
\label{eq:pseudo-Hermicitity}
    \Sigma_z H^\dag_{BdG}(k)\Sigma_z^{-1} = H_{BdG}(k),
\end{equation}
and particle-hole symmetry
\begin{equation}
\label{eq:particle-hole symmetry}
\mathcal{C} H^*_{BdG}(k) \mathcal{C}^{-1} = - H_{BdG}(-k),
\end{equation}
where $\Sigma_z= \tau_z \otimes 1_i$ and $\mathcal{C} = \tau_x \otimes 1_i$  with $1_i$ the identity matrix in the internal space and $\tau_{x,y,z}$ and $\tau_0$ the Pauli and identity matrices in Nambu space, $(\ket{+}, \ket{-})$. No symmetry conditions have been imposed on our model; pseudo-Hermiciticy (\ref{eq:pseudo-Hermicitity}) and particle-hole symmetry (\ref{eq:particle-hole symmetry}) are intrinsic to the bosonic BdG Hamiltonian.
 
We briefly mention that pseudo-Hermiticity (\ref{eq:pseudo-Hermicitity}), while not a part of  the Altland-Zirnbauer (AZ) symmetry classes \cite{altland1997PhysRevB.55.1142}, is a special case \cite{lieu2018PhysRevB.98.115135} of so-called $Q$ symmetry, one of the four fundamental Bernard-LeClair (BL) symmetry classes \cite{bernardAmdLeClair2002.207}.
The topological phases of bosonic BdG Hamiltonians are thus classified according to the BL-based 38-fold way for non-Hermitian systems \cite{kawabata2019PhysRevX.9.041015,zhou2019PhysRevB.99.235112} ---  the culmination of recent efforts \cite{esaki2011PhysRevB.84.205128,lieu2018PhysRevB.98.115135,gong2018PhysRevX.8.031079} for expanding the AZ-based 10-fold way from quadratic fermionic Hamiltonians \cite{schnyder2008PhysRevB.78.195125,kitaev2009AIP.1134.22,Ryu2010NewJournalOfPhysics.12.065010} to non-Hermitian Hamiltonians.

We assume the energy scale described by $A(k)$ to be much larger than that described by $B(k)$, which allows us to split the total Hamiltonian according to $H_{BdG}(k) = H^{(0)}(k) + H^{(1)}(k)$,
where
\begin{align}
    H^{(0)}(k)= & \mathbb{P}_+ \otimes A(k) - \mathbb{P}_- \otimes A^*(-k), \label{eq: unperturbed Hamiltonian}\\
    H^{(1)}(k)= & \tau_+\otimes B(k) - \tau_-\otimes B^*(-k), \label{eq:the perturbed hamiltonian}
\end{align}
are the unperturbed Hamiltonian and its perturbation, and 
\begin{equation}
    \mathbb{P}_\pm = (\tau_0 \pm \tau_z)/2,\qquad \tau_\pm = (\tau_x \pm i\tau_y)/2,
\end{equation}
are projection and  ladder operators in Nambu space.

A perturbative approach begins with the Schr\"{o}dinger equation of the unperturbed BdG Hamiltonian,
\begin{equation}
\label{eq:unperturbed schrodinger equation}
    H^{(0)}(k)\ket{\Psi^{(0)}(k)} = E^{(0)}(k)\ket{\Psi^{(0)}(k)},
\end{equation}
where the unperturbed eigenstate $\ket{\Psi^{(0)}}$, like any (stable) eigenstate of a BdG Hamiltonian, can be classified either as a particle state $\ket{p}$ or as a hole state $\ket{h}$, depending on whether its norm with metric $\Sigma_z$ can be scaled to $+1$ or $-1$  \cite{blaizot96QuantumTheoryBook}.

It thus  obeys the so-called bi-orthonormality relation,
\begin{equation}
\label{eq:bi-orthonormality relation}
    \bra{p} \Sigma_z \ket{p'} = + \delta_{p,p'},  \bra{h} \Sigma_z \ket{h'} = -\delta_{h,h'}, \bra{p} \Sigma_z \ket{h} =0.
\end{equation}
From now on, whenever no confusion is likely to arise, we ignore the argument $k$ to eigenenergies and eigenstates so as to reduce clutter in notations.

Let  $\ket{\psi_p}$ and  $ \ket{\psi_h}$ be the eigenstates of the Hermitian matrix $A(k)$ and $-A^*(-k)$ with eigenvalues $E^{(0)}_p$ and $E^{(0)}_h$, respectively,
\begin{equation}
\label{eq:particle and hole eigenstates}
    A(k)\ket{\psi_p} =  E^{(0)}_p\ket{\psi_p},
    -A^*(-k) \ket{\psi_h} =  E^{(0)}_h\ket{\psi_h},
\end{equation}
where $ \ket{\psi_p}$ and $\ket{\psi_h}$ obey the usual orthonormality relation, 
\begin{equation}
    \bra{\psi_p}\ket{\psi_{p'}} = \delta_{p,{p'}}, \qquad \bra{\psi_h}\ket{\psi_{h'}} = \delta_{h,{h'}}.
\end{equation}
within particle states and hole states, independently.
In terms of  $ \ket{\psi_p}$ and $\ket{\psi_h}$, the particle and hole states of Eq. (\ref{eq:unperturbed schrodinger equation}) are given as
\begin{equation}
     \ket{p} = \ket{+} \otimes \ket{\psi_p}, \qquad \ket{h} = \ket{-} \otimes \ket{\psi_h}. 
\end{equation}

Nakamura et al. \cite{nakamura2008PhysRevA.77.043601}, following earlier hints \cite{wu2001PhysRevA.64.061603,kawaguchi2004PhysRevA.70.043610}, pointed out that the appearance of complex eigenvalues is always accompanied by a level crossing, i.e. a degeneracy between a hole and particle state when in the absence of pairing interactions. In the spirit of \cite{nakamura2008PhysRevA.77.043601}, we focus on a particular particle and hole state of the unperturbed Hamiltonian,
\begin{equation}
\label{eq:p and h states}
\ket{p_0} = \ket{+} \otimes \ket{\psi_{p_0}},\qquad \ket{h_0} =  \ket{-} \otimes \ket{\psi_{h_0}},
\end{equation}
which are degenerate with energy
\begin{equation}
    E^{(0)}_{p_0} = E^{(0)}_{h_0} \equiv E_0,
\end{equation}
and we hope to gain insight into how the pairing interaction lifts this degeneracy.

We first apply the standard perturbation ansatz,
\begin{equation}
E =\sum_{n}E^{(n)}, \quad
    \ket{\Psi} = \sum_{n} \ket{\Psi^{(n)}},\quad n = 0, 1,\cdots,
\end{equation}
to the Schr\"{o}dinger equation 
\begin{equation}
    H_{BdG}(k)\ket{\Psi(k)} = E(k)\ket{\Psi(k)},
\end{equation}
where $E^{(n)}$ and $\ket{\Psi^{(n)}}$ are the $n$th order correction [in $H^{(1)}$] to the energy and corresponding eigenstate. In doing so, we arrive at a hierarchy of higher order equations,
\begin{align}
    (H^{(0)} - E_0)\ket{\Psi^{(1)}} = & (E^{(1)} - H^{(1)})\ket{\Psi^{(0)}}, \label{eq:order 1}\\
     (H^{(0)} - E_0)\ket{\Psi^{(2)}} = &  E^{(2)} \ket{\Psi^{(0)}} \nonumber \\
 & + (E^{(1)} - H^{(1)})\ket{\Psi^{(1)}}, \label{eq:order 2} \\
\vdots \nonumber
\end{align}
in addition to the zeroth order equation (\ref{eq:unperturbed schrodinger equation}).

We then apply the degenerate perturbation ansatz in which $\ket{\Psi^{(0)}}$ is assumed to be the superposition of the degenerate states,
\begin{equation}
  \ket{\Psi^{(0)}} = \alpha \ket{p_0} + \beta \ket{h_0},   
\end{equation}
 while $\ket{\Psi^{(n > 0)}}$ is assumed to be the superposition of the particle and hole states,
\begin{equation}
\label{eq:Psi n}
     \ket{\Psi^{(n)}}=\sum'_p a^{(n)}_p \ket{p} + \sum'_h a^{(n)}_h \ket{h}
\end{equation}
where the primes are to stress that the degenerate states, $\ket{p_0}$ and $\ket{h_0}$, are excluded from the first superposition over the particle states and the second superposition over the hole states, respectively.

We stress that since the unperturbed Hamiltonian (\ref{eq: unperturbed Hamiltonian}) is Hermitian, all unperturbed eigenstates, upon which we make our perturbation expansion, are stable and form a complete Hilbert space.  Further, in our derivation below, inner products are always between two unperturbed eigenstates.  As a result, even though the total Hamiltonian (with pairing interactions) is pseudo-Hermitian, issues associated with a pseudo-Hermitian system such as zero modes and modes whose norms with metric $\Sigma_z$ vanish \cite{blaizot96QuantumTheoryBook}, never arise throughout our derivation bellow.

Joining Eq. (\ref{eq:order 1}) with $\bra{p_0} \Sigma_z$ and $\bra{h_0} \Sigma_z$ (from the left),  we arrive at the eigenvalue equation for $E^{(1)}$,
\begin{equation}
\label{eq:first-order perturbation Matrix}
    \begin{pmatrix}
     \bra{p_0}\Sigma_z H^{(1)} \ket{p_0} &  \bra{p_0}\Sigma_z H^{(1)}\ket{h_0} \\
      -\bra{h_0}\Sigma_z H^{(1)} \ket{p_0} &
       -\bra{h_0}\Sigma_z H^{(1)}\ket{h_0}
    \end{pmatrix} \begin{pmatrix}
     \alpha \\
    \beta
    \end{pmatrix} = E^{(1)}\begin{pmatrix}
     \alpha \\
    \beta
    \end{pmatrix},
\end{equation}
where use of the bi-orthonormality condition (\ref{eq:bi-orthonormality relation}) has been made.  The eigenvalue $E^{(1)}$ can be complex since the matrix in Eq. (\ref{eq:first-order perturbation Matrix}) can be  non-Hermitian. Inserting Eq. (\ref{eq:the perturbed hamiltonian}) into Eq. (\ref{eq:first-order perturbation Matrix}), we simplify Eq. (\ref{eq:first-order perturbation Matrix}) into 
\begin{equation}
    \begin{pmatrix}
     0 &  \bra{\psi_{p_0}}B(k)\ket{\psi_{h_0}}  \\
      - \bra{\psi_{p_0}}B(k)\ket{\psi_{h_0}}^*  & 0
    \end{pmatrix}\begin{pmatrix}
     \alpha \\
    \beta
    \end{pmatrix} = E^{(1)}\begin{pmatrix}
     \alpha \\
    \beta
    \end{pmatrix},
\end{equation}
from which we find 
\begin{equation}
E^{(1)} = \pm i \abs{\bra{\psi_{p_0}}B(k)\ket{\psi_{h_0}}}.
\end{equation}
To write this in a more enlightening form, we introduce an unconventional commutator defined as
\begin{equation}
\label{eq:a commutator}
    \left \lceil A(k),B(k) \right\rfloor   \equiv  A(k)B(k) - B(k) A^*(-k),
\end{equation}
which is different from the conventional commutator $[A(k),B(k)]$ unless $A(k)$ is both real and even in $k$.  With the help of Eq. (\ref{eq:particle and hole eigenstates}), we find that
\begin{equation}
\label{eq:an identity associated with [A,B]}
    \bra{\psi_p}\left \lceil A(k),B(k) \right\rfloor \ket{\psi_h}
    = (E_p^{(0)} + E_h^{(0)})\bra{\psi_p}B(k)\ket{\psi_h},
\end{equation}
which holds irrespective of whether $\ket{p}$ and $\ket{h}$ are the degenerate states. For the special case of the degenerate states, Eq. (\ref{eq:an identity associated with [A,B]}) becomes
\begin{equation}\bra{\psi_{p_0}}\left \lceil A(k),B(k) \right\rfloor \ket{\psi_{h_0}}
    = 2E_0\bra{\psi_{p_0}}B(k)\ket{\psi_{h_0}},
\end{equation}
which  immediately leads to the following theorem.
\begin{theorem}
\label{theorem1}
    Let Eq. (\ref{eq:H_BdG}) be the BdG Hamiltonian for a bosonic system.  Let a pair of particle and hole states be degenerate with energy $E_0$ in the absence of $B(k)$. Then, a weak $B(k)$ lifts the degeneracy, splitting $E_0$ into a pair of complex conjugate energies, $E_0 + i |E^{(1)}|$ and $E_0 - i |E^{(1)}|$, where $\abs{E^{(1)}}$ is given by
     \begin{equation}
\label{eq:first order energy correction}
    |E^{(1)}|=  \abs{\bra{\psi_{p_0}}\left \lceil A(k),B(k) \right\rfloor  \ket{\psi_{h_0}}}/{2|E_0|},
\end{equation}
 which is valid to first order in $B(k)$, provided that
\begin{equation}
\label{E0 condition}
    |E_0| \gg \sqrt{\abs{\bra{\psi_{p_0}}\left \lceil A(k),B(k) \right\rfloor  \ket{\psi_{h_0}}}/2},
\end{equation}
where $\left \lceil A(k),B(k) \right\rfloor $ is the unconventional commutator defined in Eq. (\ref{eq:a commutator}) and 
$\ket{\psi_{p_0}}$ and  $ \ket{\psi_{h_0}}$ are the eigenstates defined in Eq. (\ref{eq:particle and hole eigenstates}).
\end{theorem}
This theorem applies to a finite bosonic system described by a BdG Hamilonian in which the number nonconserving (pairing) part is much smaller in energy scale than the number conserving part and further the number conserving part supports degenerate particle and hole states. In the following, we shall refer to the states whose energies satisfy condition (\ref{E0 condition}) as high-$\abs{E_0}$ states.

Theorem \ref{theorem1} \cite{ling2020PRL.Note1}
states that a pair of degenerate high-$\abs{E_0}$ particle and hole states are unstable against $B(k)$ if the transition between them is not $H_{ph}(k)$-forbidden, where
\begin{equation}
H_{ph}(k) \equiv \left \lceil A(k),B(k) \right\rfloor/(2|E_0|),
\end{equation}
is the effective coupling between the two degenerate states. Creating a stable BdG system amounts to developing selection rules for $H_{ph}(k)$-forbidden transitions as far as first-order perturbation theory is concerned. 

In particular, if $A(k)$ ``commutes" with $B(k)$, i.e., $\left \lceil A(k),B(k) \right\rfloor =0 $,  all transitions between high-$\abs{E_0}$ degenerate states are $H_{ph}(k)$-forbidden. This raises the question of whether complex energies can arise from higher than first-order corrections, thereby causing the degenerate states to be unstable. To address this question, we have to extend the degenerate perturbation theory to second order because the degeneracy is not lifted at first order.  We first join Eq. (\ref{eq:order 2}) with  $\bra{p_0}\Sigma_z$ and $\bra{h_0}\Sigma_z$, which leads to 
\begin{equation}
\begin{split}
\label{eq:alpha beta 2}
   \bra{p_0}\Sigma_z H^{(1)}\ket{\Psi^{(1)}}  & =  E^{(2)}\alpha, \\
   -\bra{h_0}\Sigma_z H^{(1)}\ket{\Psi^{(1)}}  & =  E^{(2)}\beta . 
\end{split}
\end{equation}
 We next employ Eq. (\ref{eq:order 1}) to express $\ket{\Psi^{(1)}}$ in terms of $ \ket{\Psi^{(0)}}$ as
\begin{equation}
\label{eq:a_p a_h 1}
\begin{split}
 a^{(1)}_{p} & = -\frac{\bra{p}\Sigma_z H^{(1)} \ket{\Psi^{(0)}}}{E_p^{(0)} - E_0} =- \frac{\bra{\psi_p} B(k) \ket{\psi_{h_0}}}{E_p^{(0)} - E_0}\beta,\\
 a^{(1)}_h & = \frac{\bra{h}\Sigma_z H^{(1)} \ket{\Psi^{(0)}}}{E_h^{(0)} - E_0} = \frac{ \bra{\psi_h} B^*(-k) \ket{\psi_{p_0}} }{E_h^{(0)} - E_0} \alpha,
\end{split}
\end{equation}
which involve inner products such as $\bra{\psi_p} B(k) \ket{\psi_{h_0}}$. It follows from Eq. (\ref{eq:an identity associated with [A,B]}) that when $ \left \lceil A(k),B(k) \right\rfloor=0$, the only chance for $\bra{\psi_p} B(k) \ket{\psi_{h}}$ not to vanish is when $E_p^{(0)}$ and  $E_h^{(0)}$ (which are the energies of states $\ket{\psi_p}$ and $\ket{\psi_h}$) are equal in magnitude but opposite in sign:
\begin{equation}
\label{eq:useful relation}
    \bra{\psi_p} B(k) \ket{\psi_{h}}= \bra{\psi_p} B(k) \ket{\psi_{h}} \delta_{E_p^{(0)}, -E_h^{(0)}}.
\end{equation}
With this,  Eq. (\ref{eq:a_p a_h 1}) can be expressed as
\begin{equation}
\label{eq: a_p and a_h 1 simplified}
\begin{split}
    a^{(1)}_{p} &= + \frac{\bra{\psi_p} B(k)\ket{\psi_{h_0}}} {2E_0}\beta,\\
 a^{(1)}_h  & = - \frac{\bra{\psi_h} B^*(-k) \ket{\psi_{p_0}} }{2 E_0} \alpha.
\end{split}
\end{equation}
Inserting this into Eq. (\ref{eq:alpha beta 2}), we find that the two equations in Eq. (\ref {eq:alpha beta 2}) decouple, leading immediately to two eigenstates: one with energy 
\begin{equation}
\label{eq: E-}
     E_{p_0}^{(2)} = \frac{-1}{2E_0}\sum'_h \bra{\psi_{p_0}} B(k) \ket{\psi_h}\bra{\psi_{h}} B^*(-k) \ket{\psi_{p_0}}
\end{equation}
and $\alpha = 1 $ and $\beta = 0$
and the other with energy
\begin{equation}
\label{eq: E+}
     E_{h_0}^{(2)} = \frac{-1}{2E_0}\sum'_p \bra{\psi_{h_0}} B^*(-k) \ket{\psi_p}\bra{\psi_{p}} B(k) \ket{\psi_{h_0}}
\end{equation}
and $\alpha = 0 $ and $\beta = 1$. We can remove the prime from each sum by virtue of Eq. (\ref{eq:useful relation}).  Finally, we apply, independently, the closure identities for $\ket{\psi_p}$ and $\ket{\psi_h}$ to cast $E_{p_0}^{(2)}$ and $E_{h_0}^{(2)}$ into the simpler forms,
\begin{equation}
\label{eq:second-order energy splitting}
\begin{split}
    E_{p_0}^{(2)} &=- \frac{\bra{\psi_{p_0}} B(k) B^*(-k) \ket{\psi_{p_0}}}{2E_0},\\
     E_{h_0}^{(2)} &=- \frac{\bra{\psi_{h_0}} B^*(-k) B(k) \ket{\psi_{h_0}}}{2E_0},
\end{split}
\end{equation}
which are real since they are the averages of two Hermitian operators, $B(k) B^*(-k)$ and $B^*(-k) B(k)$.  The Hermiticity of these two operators follows from  $B(k) = B^T(-k)$ in Eq. (\ref{eq:A and B Properties}), a consequence of  $\Sigma_z H_\mathrm{BdG}(k)$ being inherently Hermitian. As we explain in Appendix \ref{sec:Higher than second-order energy corrections}, this same property guarantees that the energy correction be real at any higher order, as long as $\left \lceil A(k),B(k) \right\rfloor =0$.

It then follows that a sufficient condition for creating a stable high-$\abs{E_0}$ state is that $A(k)$ ``commutes" with $B(k)$, i.e. the unconventional commutator (\ref{eq:a commutator}) vanishes. This may be further understood by looking at the square of $H_{BdG}$, which can be formulated in terms of $\left \lceil A,B \right\rfloor$ as
\begin{equation}
\label{eq:H^2}
\begin{split}
    H^2_{BdG}(k) = &\begin{pmatrix}
     A_2(k)&  \left \lceil A(k),B(k) \right\rfloor \\
     -\left \lceil A(k),B(k) \right\rfloor^\dag & A_2^*(-k)
    \end{pmatrix},
\end{split}
\end{equation}
where 
\begin{equation}
A_2(k) \equiv A^2(k) - B(k)B^*(-k)
\end{equation}
is a Hermitian operator. From Eq. (\ref{eq:H^2}) we see that only when $A(k)$ and $B(k)$ ``commute" is $ H^2_{BdG}(k)$ Hermitian.  Thus, it is the condition  $\left \lceil A(k),B(k) \right\rfloor=0$ [or equivalently $H_{ph}(k)=0$] that underlies the suggestion that systems with Hermitian $H^2_{BdG}(k)$ may be made to operate in a manner free of bulk instabilities \cite{galilo15PhysRevLett.115.245302}.

\section{Applications: A B\lowercase{d}G extension of spinful Haldane model}
\label{sec:application}

In this section, we apply the theorem to a generalization of a model proposed by Galilo et al. \cite{galilo15PhysRevLett.115.245302} for creating a topological atom amplifier in spin-1 cold atoms in a honeycomb lattice through a quench process. In Appendix \ref{appendix 1}, we give additional details for the (postquench) quadratic Hamiltonian describing quantum fluctuations about the initial state where all atoms are condensed to the spin-0 component. Just as in their model, the total Hamiltonian is divided into two independent sectors, one for the spin-0 component  and the other for the spin-$\pm 1$ components.  The spin-$0$ component is stable and not affected by the quench.  The sector for spin-$\pm 1$ components is a BdG extension of a spinful Haldane model and evolves in time under the quench.

We thus focus on the spin-$\pm 1$ sector, a pseudo spin-$1/2$ system, on a lattice stripe geometry with open boundaries along $y$ and periodic (zigzag) boundaries along $x$.  The internal space is the tensor product $\mathscr{H}_I\otimes \mathscr{H}_s \otimes \mathscr{H}_\sigma$, where $\mathscr{H}_I= (\ket{1},\ket{2},\cdots, \ket{N_y})$ is the unit cell space with $N_y$ the number of $y$ unit cells, $\mathscr{H}_s=(\ket{\uparrow},\ket{\downarrow})$ is the spin-$1/2$ space, and $\mathscr{H}_\sigma=(\ket{A},\ket{B})$ is the sublattice space. We describe the system using the bosonic BdG Hamiltonian (\ref{eq:H_BdG}) with 
\begin{equation}
\label{eq:A and B for our model}
\begin{split}
   A(k) = & I_- \otimes \alpha(k) + I_0 \otimes\beta(k) + I_+\otimes \gamma(k),\\
   B = & I_0\otimes \xi,
\end{split}
\end{equation}
where $ A(k)$ is a tridiagonal matrix (real and symmetric) with $k$ being the momentum vector along $x$, $B$ is a diagonal matrix (real and $k$-independent), and $I_0$, $I_-$ and $I_+$ are the main-, sub- and super-diagonal identity matrices in $\mathscr{H}_I$. 
In Eq. (\ref{eq:A and B for our model}), 
\begin{equation}
\label{eq: alpha beta gamma stripe}
\begin{split}
   \alpha(k) = & 2t_2\cos\left( \phi + \frac{k}{2} \right) s_z\otimes \sigma_z - t_1s_0\otimes\sigma_+, \\
   \beta(k) = &  2t_2\cos\left( \phi - k \right) s_z\otimes\sigma_z +m s_0\otimes\sigma_z \\
   &+ q s_0\otimes\sigma_0 - 2 t_1\cos\frac{k}{2}s_0\otimes \sigma_x, \\
   \gamma(k) =&2t_2\cos\left( \phi + \frac{k}{2} \right)  s_z\otimes\sigma_z - t_1s_0\otimes\sigma_-,\\
   \xi = & n_B c_2  s_x \otimes\sigma_0,
\end{split}
\end{equation}
are all real $4\times 4$ matrices in $\mathscr{H}_s \otimes \mathscr{H}_\sigma$, where $s_{x,y,z}$ ($s_0$) and  $\sigma_{x,y,z} $ ($\sigma_0$) are the Pauli (identity) matrices in $\mathscr{H}_s$  and $\mathscr{H}_\sigma$ and $\sigma_\pm = (\sigma_x \pm i \sigma_y)/2$. Here, $t_1$ is the nearest-neighbor hopping amplitude, $t_2e^{\pm i\phi}$ is the next-nearest-neighbor hopping amplitude introduced by Haldane \cite{haldane88PhysRevLett.61.2015}, where $\pm$ alternates periodically in the manner of Kane and Mele \cite{kane2005PhysRevLett.95.146802}, $q$ and $m$ measure, respectively, the nonstaggered and staggered onsite potential, $n_B$ is the filling factor for the condensed atoms, and $c_2$ is the two-body spin interaction strength \cite{ho1998PhysRevLett.81.742,ohmi1998doi:10.1143/JPSJ.67.1822}.    

\begin{figure}[t]
	\centering
	\includegraphics[width=0.45\textwidth]{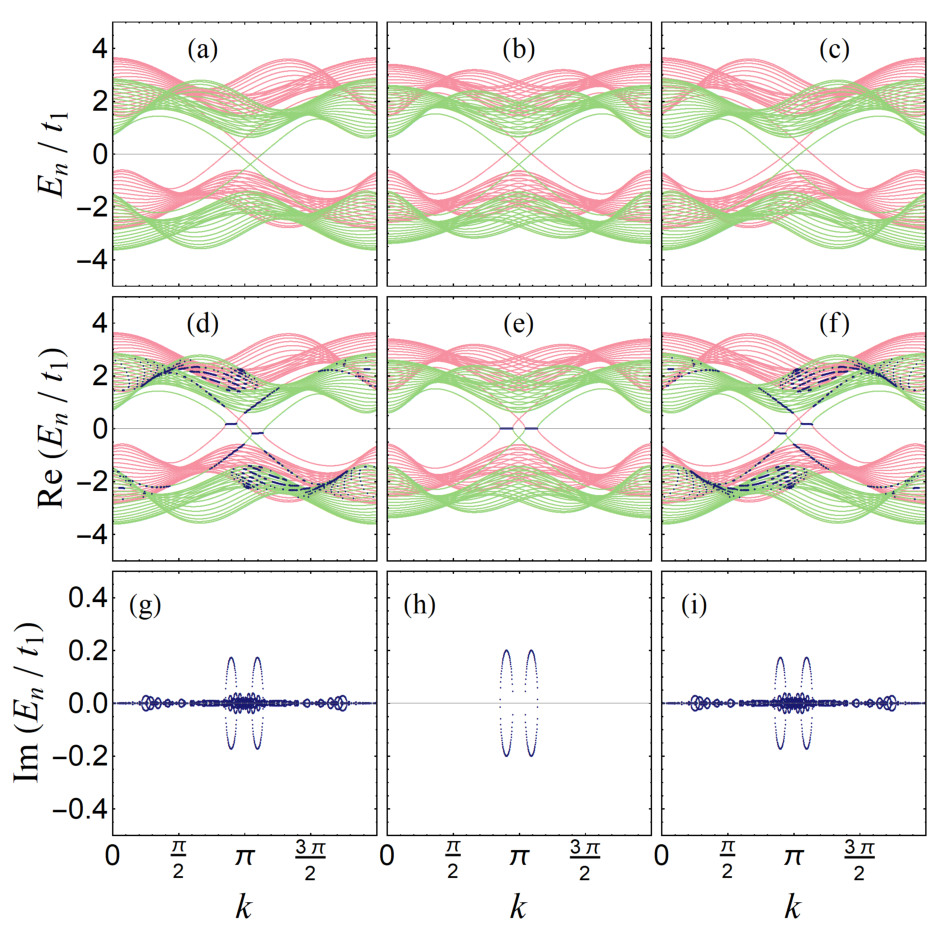}
	\caption{ Excitation energy spectrum, $E_n(k)$ in the $\Uparrow$ sector [the eigenvalue of $H_{\Uparrow}$ in Eq. (\ref{eq:H Uparrow and Downarrow})] for $m = 0$, $ n_Bc_2 =  0.2 t_1$ and $\phi =$ $\frac{\pi}{2} - \frac{\pi}{8}$ (left column), $\frac{\pi}{2}$ (middle column), and $\frac{\pi}{2} + \frac{\pi}{8}$ (right column).  The middle row shows the real part and the bottom row shows the imaginary part.  The top row displays the spectrum in the absence  of pairing interactions.  Additional parameters are $t_1=1,t_2=0.54 t_1, q=0.4t_1$ and $N_y =20$. Red, green, and dark blue indicate positive norm (particle), negative norm (hole), and zero norm (unstable) states, respectively.
	}
	\label{fig:fig1}
\end{figure}
\begin{figure}[t]
	\centering
	\includegraphics[width=0.45\textwidth]{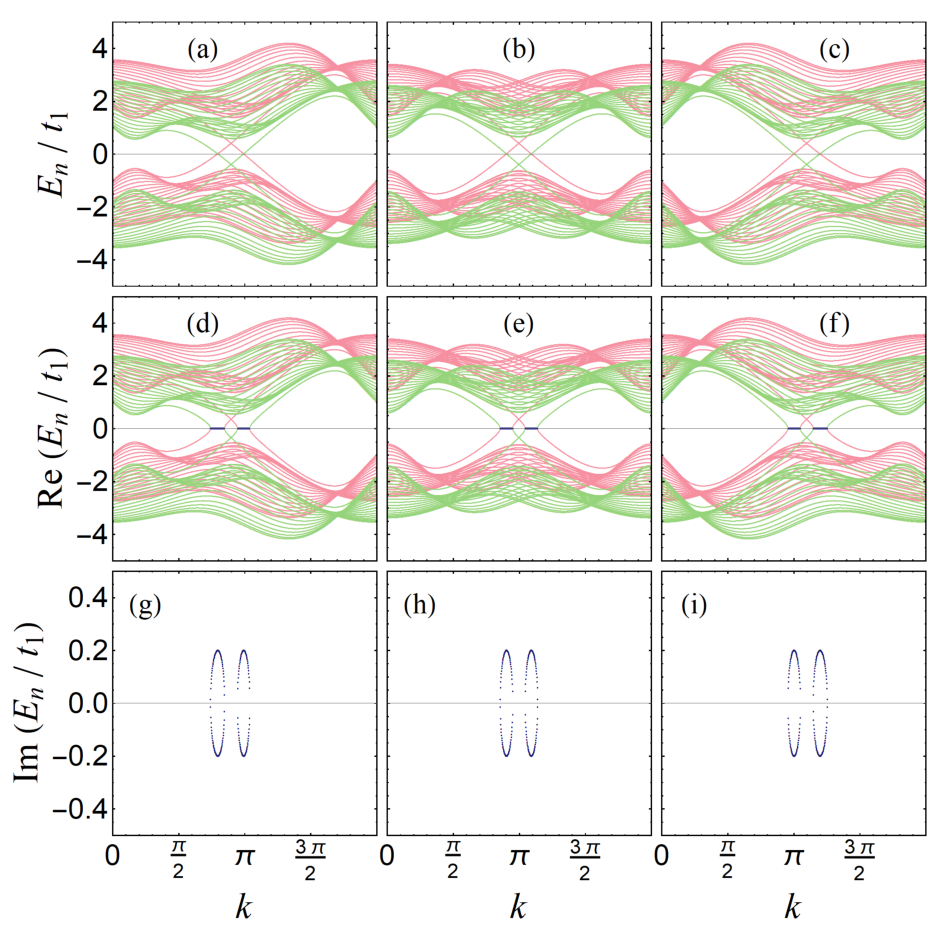}
	\caption{ Excitation energy spetrum, 
	 $E_n(k)$ in the $\Uparrow$ sector for $\phi=\frac{\pi}{2}$, $n_Bc_2 = 0.2t_1$ and $m =-t_1$ (left column), $0$ (middle column), and $+t_1$ (right column).  The middle row shows the real part and the bottom row shows the imaginary part. The top row displays the spectrum in the absence of pairing interactions.
		The remaining parameters and the color scheme are the same as in Fig. \ref{fig:fig1}.
	}
	\label{fig:fig2} 
\end{figure}

 $H_{BdG}(k)$ is neither time reversal nor inversion symmetric since we allow the phase $\phi$  and the staggered potential amplitude $m$ to be arbitrary where the time-reversal operator is defined as $\mathcal{T} = \tau_0 \otimes I_0 \otimes s_x \otimes \sigma_0 K$ with $K$ being complex conjugation and the inversion symmetry operator is defined as $\mathcal{P} = \sigma_x$ (which is short for $ \tau_0 \otimes I_0 \otimes s_0 \otimes \sigma_x $).  This is different than in \cite{galilo15PhysRevLett.115.245302}, where $\phi$  and $m$ are fixed to $\pi/2$ and $0$, respectively.  Nor is $H_{BdG}(k)$ $s_z$-rotation invariant since although $A(k)$ conserves $s_z$, $B$ does not. Instead, it transforms under $s_z$ (which is short for $\tau_0\otimes I_0 \otimes s_z\otimes \sigma_0$) in the manner of pseudo-Hermiticity,
 \begin{equation}
 \label{sz symmetry}
    s_z H_{BdG}(k)s_z^{-1} = H^\dag_{BdG}(k),
\end{equation}
due to spin and momentum conservation during a collision.  Equation (\ref{sz symmetry}), together with the generic pseudo-Hermiticity in Eq. (\ref{eq:pseudo-Hermicitity}), implies the existence of a unitary symmetry, $[J_z, H_{BdG}(k)]=0$, where $ J_z \equiv \tau_z \otimes I_0 \otimes s_z \otimes \sigma_0$ is the $z$-component of the spin rotation generator represented on Nambu (particle-hole) space \cite{altland1997PhysRevB.55.1142}.  
Being diagonal, $J_z$ partitions $H_{BdG}(k)$ into
\begin{equation}
    H_{BdG}(k)=H_\Uparrow(k) \oplus H_\Downarrow(k),
\end{equation}
where
\begin{equation}
\label{eq:H Uparrow and Downarrow}
    H_{\Uparrow/\Downarrow}(k) = 
    \begin{pmatrix}
    A_{\uparrow/\downarrow}(k) & c_2n_B I_0\otimes \sigma_0\\
    - c_2n_B I_0 \otimes \sigma_0 & -A_{\downarrow/\uparrow}(-k)
\end{pmatrix}
\end{equation}
are the Hamiltonians in the degenerate subspaces of $J_z$ in which $J_z$ has eigenvalue $+1$ and $-1$, respectively.
$A_{\uparrow/\downarrow}(k)$ are matrices in $\mathscr{H}_I \otimes \mathscr{H}_\sigma$ that are constructed as follows. They take the form of $A(k)$ in Eq. (\ref{eq:A and B for our model}), where $\alpha,\beta$, and $\gamma$ are $2\times2$ matrices that are just as in  Eq. (\ref{eq: alpha beta gamma stripe}), except $s_0$ is replaced with $+1$ and $s_z$ is replaced with $+1$ for $A_\uparrow(k)$ and with $-1$ for $A_\downarrow(k)$.

Without loss of generality, we limit our study to the $\Uparrow$ sector. We focus on systems that have edge modes and band gaps near zero energy when $B=0$ (as shown in the top row of Figs. \ref{fig:fig1} and \ref{fig:fig2}), so that the first-order perturbation theory may be applied to bulk states but not to edge modes near zero energy.
We display the real (middle row) and imaginary (bottom row) eigenvalues of $H_\Uparrow$ when $B \neq 0$ in Fig. \ref{fig:fig1} where $m$ is fixed to 0 but $\phi$ is allowed to vary and in Fig. \ref{fig:fig2} where $\phi$ is fixed to $\pi/2$ but $m$ is allowed to vary. 

To gain insight, we compute the ``commutator" in Eq. (\ref{eq:a commutator}), using Eqs. (\ref{eq:A and B for our model}) and (\ref{eq: alpha beta gamma stripe}), to find
\begin{equation}
\label{eq:unconventional commutator [A,B]}
\begin{split}
   \left \lceil  A,B \right\rfloor = &  i  4t_2 n_Bc_2  \cos\phi s_y\otimes\sigma_z \otimes\\ 
   & \left[\cos\frac{k}{2}(I_+ + I_-) + \cos k I_0 \right].
 \end{split}
\end{equation}
This result shows that only when $\phi=\pi/2$ (or $3\pi/2$) and thus only when the system has time reversal symmetry, are bulk states stable in the presence of a paring term.  This is evident in Fig. \ref{fig:fig1} where the bulk particle (red) and hole (green) energies that are degenerate  when $B = 0$  (top row)  become complex when $B \neq 0$ and  $\phi$ takes values different from $\pi/2$ (left and right columns).  Note that Eq. (\ref{eq:unconventional commutator [A,B]}) is independent of the staggered potential $m$. Thus, as long as it remains time-reversal-invariant, our system has a stable bulk even when it is no longer inversion-symmetric.
Indeed, as Fig. \ref{fig:fig2} illustrates, the presence of $m$ does not affect bulk stability; all bulk spectra are real.

We now shed some light on why our system, like the one presented by Galilo et al. \cite{galilo15PhysRevLett.115.245302}, is topological.  As mentioned earlier, because our system conserves $J_z$, $J_z$ partitions our system into two subsystems and our system inherits the topological properties of these two subsystems.  Since the subsystems in our model are built upon the Haldane model, they inherit the Dirac points and $C_3$ point group symmetry, which are unaffected by the pairing interaction [the last line in Eq.(\ref{eq:total H in position})].  The $C_3$ point group symmetry guarantees that a gap closing and re-opening transition occurs only at Dirac points \cite{bernevig2013} (see, for example, \cite{furukawa2017NewJournalOfPhysics.17.115014}).   When the spectrum for an open system changes from one without edge modes inside the bulk gaps (not shown) to one with edge modes inside the bulk gaps (which is shown in Fig. \ref{fig:fig2}), it signals that the corresponding periodic system makes a transition from a trivial gapped phase, via a gapless point at which the adiabatic condition breaks down, to a nontrivial gapped phase.  This nontrivial gapped phase is a topological state with a nontrivial Chern number (i.e. it is a Chern insulator) since it is associated with a gap closing, which is a source of Berry curvature and hence nontrivial Chern number.  In summary, our system is topological and is classified as two copies of Chern insulators.  We leave as future work a detailed and quantitative study of the topological properties of our system, including how to identify its symmetry class within the framework of 38-fold way and how to determine its bulk topological variant (the Chern number) within the context of non-Hermitian physics \cite{kawabata2019PhysRevX.9.041015}.

To study $\phi=\pi/2$, where time reversal symmetry necessitates $A_\uparrow(k) = A_\downarrow(-k)$, 
we move to the basis $\{\ket{\omega_n} \}$ \cite{galilo15PhysRevLett.115.245302}, where $\ket{\omega_n}$ is an eigenstate of $A_\uparrow(k)$,
\begin{equation}
    A_\uparrow(k)\ket{\omega_n} = \omega_n\ket{\omega_n}.
\end{equation} 
In this basis, we can take advantage $A_\uparrow(k) = A_\downarrow(-k)$ and that pairing terms in Eq. (\ref{eq:H Uparrow and Downarrow}) are proportional to the identity matrix $I_0 \otimes \sigma_0$, decomposing $H_\Uparrow(k)$ into a direct sum, $H_\Uparrow(k)= \bigoplus_n H_{\Uparrow,n}(k)$, where
\begin{equation}
    H_{\Uparrow,n}(k) = \begin{pmatrix}
    \omega_n(k) & n_Bc_2\\
    -n_Bc_2 & -\omega_n(k)
    \end{pmatrix},
\end{equation}
 is a two-state Hamiltonian.   We emphasize that this simplification holds regardless of the inversion symmetry.  The eigenenergies in the $\Uparrow$ sector are now easily found to be \begin{equation}
\label{eq:E_n(k_x)}
    E_n (k) = \pm \sqrt{\omega_n^2(k) - n_B^2c_2^2}.
\end{equation} 
As anticipated by Theorem \ref{theorem1},  a Taylor series of $E_n(k)$ does not contain the term linear in the small parameter, $n_Bc_2/\abs{\omega_n}$, and $E_n(k)$ is real except when $\abs{\omega_n(k)}<n_B c_2$, where perturbation theory breaks down. 
Equation (\ref{eq:E_n(k_x)}) also applies to edge modes. To a good approximation, we can estimate edge mode dispersions using Eq. (\ref{eq:E_n(k_x)}) with $\omega_n(k)$ given by
\begin{equation}
\label{eq:omega_n edge}
    \omega_{\mathrm{edge}}(k)= q \mp t_1\frac{ 6t_2\sin k +m}{\sqrt{t_1^2 + 16 t_2^2 \sin^2 \frac{k}{2}}},
\end{equation}
which are the edge mode dispersions of $A_{\uparrow}(k)$, the Hamiltonian of Haldane model (with $\phi=\pi/2$), for a lower half and upper half semi-infinite plane. Equation (\ref{eq:omega_n edge}), which generalizes the one for $
m = 0$ in \cite{galilo15PhysRevLett.115.245302}, indicates that $m$ constitutes another knob for selecting the momentum $k$ at which an edge mode is made to lase.

Note that we have checked for finite-size, e.g. $N_y = 20$ is used to make Figs. \ref{fig:fig1} and \ref{fig:fig2} and is large enough that boundary states come in pairs localized to opposite ends. At the top edge (of the stripe lattice with open boundaries along $y$), the particle edge state moves to the right (the positive $x$ direction) while the hole edge state moves to the left. At the bottom edge, the particle edge state moves to the left while the hole edge state moves to the right.

 Importantly, the staggered potential does not affect the bulk band stability in our model.  We stress that the reason for this is that the staggered potential, $ A_{sp} \equiv m I_0 \otimes s_0 \otimes \sigma_z$ in $A(k)$ commutes with $B =n_B c_2 I_0 \otimes s_x \otimes \sigma_0$.   If we included in $B$ a perturbation of the form $B' \propto I_0 \otimes s_x \otimes \sigma_z$ which does not commute with the nearest-neighbor tunneling terms in $A(k)$, the ``commutator" for the perturbed system with $\phi= \pi/2$ would become 
\begin{equation}
\begin{split}
   \left \lceil A,B+ B' \right\rfloor & \propto  2i t_1  \cos\frac{k}{2} I_0 \otimes s_x \otimes \sigma_y\\
   & +2 t_1 I_- \otimes s_x \otimes \sigma_+ -  2 t_1 I_+ \otimes s_x \otimes \sigma_- 
\end{split}
\end{equation}
and would thus not vanish even when the system is time-reversal-invariant.  Even though $B'$ and $A_{sp}$ preserve both time and inversion symmetries, it is $B'$ that affects bulk stability, and not $A_{sp}$, demonstrating that  the ``commutator" in Eq. (\ref{eq:a commutator}) plays a fundamental role in the quest of topological amplifiers in BdG systems. Said another way, it is $\left \lceil A,B \right\rfloor =0$ that selects the symmetries a system must possess so that its bulk states can be made stable against (weak) pairing interactions.

\section{conclusion}
\label{sec:conclusion}
In this work, we presented a theorem which quantifies how a weak pairing interaction lifts the degeneracy of particle and hole states with energy far from zero in a bosonic BdG system.  We expressed the energy splitting, which is imaginary, in terms of the unconventional commutator in Eq. (\ref{eq:a commutator}). We found that when $\left \lceil A(k),B(k) \right\rfloor=0$, the property, $B(k)=B^T(-k)$, which is inherent of the BdG Hamiltonian guarantees the energy correction to be real at any higher order.  We were thus led to treat the vanishing of this ``commutator" as a practical criterion for testing and designing a topological amplifier in a BdG system. We also studied a generalization of the model from  Galilo et al.  \cite{galilo15PhysRevLett.115.245302}, finding that as long as there is time reversal symmetry, the model can be made to act like a topological amplifier, independent of whether there is an inversion symmetry, i.e. independent of whether there exists an onsite staggered potential.

The theorem we developed is fairly general. Although we applied it only to a cold atom model in this work,  we expect it to find applications in a broad array of systems across different disciplines \cite{cooper2019RevModPhys.91.015005,ozawa2019RevModPhys.91.015006,zhang2018CommunPhys.1.97,kondo2020,gong2018PhysRevX.8.031079}.

\appendix

\section{Higher than second-order energy corrections when $A(k)$ and $B(k)$ ``commute" }
\label{sec:Higher than second-order energy corrections}

In Sec. \ref{sec:a theorem}, we focused on a particle and a hole state that are degenerate in the absence of the pairing interaction and found that when $A(k)$ and $B(k)$ ``commute", the pairing interaction, while does not contribute to the energy at first order, causes the energy to shift at second order from $E_0$ to $E_{p_0}^{(2)}$ and $E_{h_0}^{(2)}$ given by Eq. (\ref{eq:second-order energy splitting}).

In this appendix, we pursue higher order corrections with the following perturbation equations,
\begin{widetext}
\begin{align}
    (H^{(0)} - E^{(0)})\ket{\Psi^{(3)}} &= (E^{(1)} - H^{(1)})\ket{\Psi^{(2)}} + E^{(2)} \ket{\Psi^{(1)}} + E^{(3)} \ket{\Psi^{(0)}}, \label{eq:order 3}\\
    (H^{(0)} - E^{(0)})\ket{\Psi^{(4)}} &= (E^{(1)} - H^{(1)})\ket{\Psi^{(3)}} + E^{(2)} \ket{\Psi^{(2)}} + E^{(3)} \ket{\Psi^{(1)}} + E^{(4)} \ket{\Psi^{(0)}}, \label{eq:order 4}\\
\vdots \nonumber
\end{align}
\end{widetext}
   
We begin with the third-order correction  which depends not only on the zeroth- and first-order states, which we obtained in the main text, but also on the second-order state, which we now determine from Eq.  (\ref{eq:order 2}). From Eq.  (\ref{eq:order 2}),
\begin{equation}
\begin{split}
    a_p^{(2)} & = \frac{\bra{p}\Sigma_z  H^{(1)}\ket{\Psi^{(1)}}}{E_0-E_p^{(0)}}=-\frac{\sum'_h a_h^{(1)}\bra{\psi_p}B(k)\ket{\psi_h}}{E_p^{(0)} - E_0},\\
    a_h^{(2)} & = \frac{\bra{h}\Sigma_z  H^{(1)}\ket{\Psi^{(1)}}}{E_h^{(0)}-E_0}
    =\frac{\sum'_p a_p^{(1)}\bra{\psi_h}B^*(-k)\ket{\psi_p}}{E_h^{(0)} - E_0},
\end{split}
\end{equation}
which becomes, after applying Eq. (\ref{eq: a_p and a_h 1 simplified}),
\begin{equation}
\label{eq:a_p a_h 2 appendix}
\begin{split}
    a_p^{(2)} &
    =\sum'_h \frac{ \bra{\psi_p}B(k) \ket{\psi_h}\bra{\psi_h} B^*(-k) \ket{\psi_{p_0}} }{2E_0(E_p^{(0)} - E_0)} \alpha,
    \\
    a_h^{(2)} &  =\sum'_p  
    \frac{\bra{\psi_h}B^*(-k)\ket{\psi_p} \bra{\psi_p} B(k) \ket{\psi_{h_0}}}{2E_0(E_h^{(0)} - E_0)} \beta.
\end{split}
\end{equation}
In lieu of the restrictions by Eq. (\ref{eq:useful relation}), the only possible contribution to the sum in $a_p^{(2)}$ [$a_h^{(2)}$] in Eq. (\ref{eq:a_p a_h 2 appendix}) comes from the hole (particle) state in which $ E_h^{(0)} =-E_p^{(0)} = -E_0$ [$E_p^{(0)} = - E_h^{(0)} = -E_0$]. This means that all $a_p^{(2)}$ [$a_h^{(2)}$] vanish except the one associated with the degenerate particle (hole) state $\ket{p=p_0}$ ($\ket{h=h_0}$) where $E_p^{(0)}=E_0$ [$E_h^{(0)} = E_0$]. However, it is precisely $a_p^{(2)}$ and $a_h^{(2)}$, which are associated with the degenerate states, that are excluded from  Eq. (\ref{eq:Psi n}) for the second-order correction.  We conclude that 
\begin{equation}
    a_p^{(2)}=a_h^{(2)} = 0.
\end{equation}

The next step depends on whether $E_{p_0}^{(2)}$ equals $E_{h_0}^{(2)}$, that is, whether the second order solution is degenerate or nondegenerate. In the degenerate case, the methods presented in Sec.  \ref{sec:a theorem} can be generalized to higher than second order. For this reason, we do not present them here.  We instead present the nondegenerate case. Though we do not show it, it can be shown that the final results for the degenerate and nondegenerate cases are the same.

Let us briefly summarize the results through second order when $\left \lceil A(k),B(k) \right\rfloor=0$. At second order and when $E_{p_0}^{(2)} \neq E_{h_0}^{(2)}$, the pairing interaction splits the degenerate states, $\ket{p_0}$ and $\ket{h_0}$, with energy $E_0$ into two states. The first state has $\alpha=1,\beta=0$, and
\begin{equation}
\label{eq:E2- branch}
\begin{split}
    E^{(0)} = E_0, & \quad \ket{\Psi^{(0)}} = \ket{p_0}, \\
E^{(1)} = 0, & \quad \ket{\Psi^{(1)}} = \sum'_h a_h^{(1)} \ket{h},  \\
E^{(2)} = E^{(2)}_{p_0}, & \quad \ket{\Psi^{(2)}} = 0,
\end{split}
\end{equation}
where $E_{p_0}^{(2)}$ is given by Eq. (\ref{eq:second-order energy splitting}) and
\begin{equation}
\label{eq: a_h 1 alpha}
    a^{(1)}_h  =  \frac{\bra{\psi_h} B^*(-k) \ket{\psi_{p_0}} }{2 E_0},
\end{equation}
which follows from Eq. (\ref{eq: a_p and a_h 1 simplified}) with $\alpha=1$.  The second sate has $\alpha = 0,\beta = 1$ and
\begin{equation}
\label{eq:E2+ branch}
\begin{split}
    E^{(0)}  =E_0, & \quad \ket{\Psi^{(0)}} = \ket{h_0}, \\
E^{(1)} = 0, & \quad \ket{\Psi^{(1)}} = \sum'_p a_p^{(1)} \ket{p}, \\
E^{(2)} = E^{(2)}_{h_0}, & \quad \ket{\Psi^{(2)}} = 0,
\end{split}
\end{equation}
where $E_{h_0}^{(2)}$ is given by Eq. (\ref{eq:second-order energy splitting}) and
\begin{equation}
     a^{(1)}_{p} = - \frac{\bra{\psi_p} B(k)\ket{\psi_{h_0}}} {2E_0}.
\end{equation}
which follows from Eq. (\ref{eq: a_p and a_h 1 simplified}) with $\beta = 1$.

Under our assumption that degeneracy is lifted at second order, corrections higher than second order are computed with nondegenerate perturbation theory. Without loss of generality, we focus on the $E_{p_0}^{(2)}$ branch. We find easily from Eq. (\ref{eq:order 3}), with the help of Eq. (\ref{eq:E2- branch}), that
\begin{equation}
    E^{(3)}\equiv E^{(3)}_{p_0} = 0, \qquad \ket{\Psi^{(3)}} = \sum'_h a_h^{(3)}\ket{h},
\end{equation}
where 
\begin{equation}
\label{eq:a_p 3}
\begin{split}
    a_h^{(3)} &= - \frac{E^{(2)}_{p_0}a_h^{(1)}}{E_h^{(0)}-E_0}= E^{(2)}_{p_0}\frac{\bra{\psi_h} B^*(-k) \ket{\psi_{p_0}} }{(2 E_0)^2}
\end{split}
\end{equation}
with the last equality following from Eq. (\ref{eq:useful relation}) and Eq. (\ref{eq: a_h 1 alpha}).   From Eq. (\ref{eq:order 4}), we then have the fourth-order correction to the energy,
\begin{equation}
    E^{(4)} =E^{(4)}_{p_0} \equiv -\bra{p_0}\Sigma_z H^{(1)} \ket{\Psi^{(3)}}
\end{equation}
which, when use of Eq. (\ref{eq:a_p 3}) is made, can be written as
\begin{equation}
\begin{split}
    E^{(4)}_{p_0} &= E^{(2)}_{p_0}\frac{1}{2E_0} \times\\
& \frac{(-1)}{2E_0}\sum'_h \bra{\psi_{p_0}} B(k) \ket{\psi_h}  \bra{\psi_h} B^*(-k) \ket{\psi_{p_0} }.
\end{split}
\end{equation}
Compared to Eq. (\ref{eq: E-}), we find that the second line is simply $E^{(2)}_{p_0}$, and thus
\begin{equation}
    E^{(4)}_{p_0} = \frac{\left[E^{(2)}_{p_0}\right]^2}{2E_0}.
\end{equation}
Since $E^{(2)}_{p_0}$ and $E_0$ are both real, so too is $E^{(4)}_{p_0}$. 

Following the same steps that begin from Eq. (\ref{eq:E2+ branch}), we easily arrive at
\begin{equation}
  E^{(3)} \equiv E^{(3)}_{h_0}=0,\quad E^{(4)}\equiv E^{(4)}_{h_0} = \frac{\left[E^{(2)}_{h_0}\right]^2}{2E_0},
\end{equation}
for the $E_{h_0}^{(2)}$ branch.

Generalizing to higher orders is straightforward.  Upon doing so, one finds that all odd higher-order energy corrections vanish while all higher-order even ones can be expressed in terms of $E^{(2)}$ and $E_0$ and are therefore real. We note that this same conclusion is found if we had instead assumed that the second order corrections are degenerate. 

We end this appendix by stressing that a bosonic BdG Hamiltonian $H_\mathrm{BdG}(k)$ has an intrinsic property that $\Sigma_z H_\mathrm{BdG}(k)$ is Hermitian, which ensures $B(k) = B^T(-k)$.  And when $\left \lceil A(k),B(k) \right\rfloor =0$, it is $B(k) = B^T(-k)$ that guarantees the energy correction to be real at any order in perturbation theory.

\section{Hamiltonian for a BdG extension of a spinful Haldane model}

\label{appendix 1}



In the main text, we study a generalization of a model proposed by Galilo et al. \cite{galilo15PhysRevLett.115.245302}.  In this Appendix, we provide details of the Hamiltonian for this generalization.  Motivated by the experimental realization of the Haldane model \cite{haldane88PhysRevLett.61.2015} by the Esslinger group \cite{jotzu2014Nature.515.237} in ultracold atoms in honeycomb optical lattices, Galilo et al. \cite{galilo15PhysRevLett.115.245302} proposed a spin-1 extension of such a system where the spin-orbit coupling is proportional to the spin projection along $z$, $S_z$, where $S_{x,y,z}$ are the spin components of the spin-1 vector $\vb{S}$. 
The generalization we study has a Hamiltonian consisting of three parts.  
First, the hopping Hamiltonian
\begin{equation}
	\label{eq:H_hop}
	\hat{H}_{hop} = -t_1 \sum_{\left< \vb{ij} \right>} \hat{b}'^\dag_{\vb{i}} \hat{b}'_{\vb{j}} + t_2\sum_{\left<\left< \vb{ij} \right> \right>} e^{-i\nu_{\vb{ij}} \phi} \hat{b}'^\dag_{\vb{i}} S_z \hat{b}'_{\vb{j}}.
\end{equation}
where $\hat{b}'_{\vb{i}}=(\hat{b}'_{\vb{i},+1},\hat{b}'_{\vb{i},0}, \hat{b}'_{\vb{i},-1})$ is the field operator,  $\hat{b}'_{\vb{i},M}$ is the annihilation operator of a boson with spin-$M$ component at site $\vb{i}$,  $t_1$ is the nearest-neighbor hopping amplitude, and $t_2e^{i\nu_{\vb{ij}} \phi}$ is the next-nearest-neighbor hopping amplitude introduced by Haldane \cite{haldane88PhysRevLett.61.2015}, where $\nu_{\vb{ij}}$ alternates between $+1$ and $-1$ periodically in the manner of Kane and Mele \cite{kane2005PhysRevLett.95.146802}.
Second, the onsite two-body interaction that preserves the spin rotation invariance \cite{ho1998PhysRevLett.81.742,ohmi1998doi:10.1143/JPSJ.67.1822},
\begin{equation}\hat{H}_{col} = \frac{c_0}{2}\sum_{\vb{i}}(\hat{b}'^\dag_{\vb{i}} \hat{b}'_{\vb{i}})^2 +\frac{c_2}{2}\sum_{\vb{i}}( \hat{b}'^\dag_{\vb{i}} \vb{S} \hat{b}'_{\vb{i}})^2,
\end{equation}
where $c_0$ and $c_2$ represent, respectively, the density and spin interaction strength. The onsite collision is the source of pairing interactions,  which is an essential ingredient of a BdG system. 
Third, the onsite potential (with staggering) 
\begin{equation}
	\hat{H}_{pot} = q' \sum_{\vb{i}} \hat{b}'^\dag_{\vb{i}} S_z^2 \hat{b}'_{\vb{i}} 
	+ m \sum_{\vb{i}} \xi_{\vb{i}}  \hat{b}'^\dag_{\vb{i}} S_z^2 \hat{b}'_{\vb{i}},
\end{equation}
which describes the quadratic Zeeman shifts for spinor condensates which can be generated by external magnetic fields \cite{stamperKurn2013RevModPhys.85.1191} or by microwave fields  \cite{gerbier2006PhysRevA.73.041602} where $\xi_{\vb{i}}$ is $+1$ for sites on sublattice $A$ and $-1$ for sites on sublattice $B$.

As in \cite{galilo15PhysRevLett.115.245302}, the onsite potential initially supports a polar condensate where all atoms are condensed to the $\ket{S=1,M=0}$ spin mode and are assumed to be uniformly distributed in space. We then apply a quench process, which abruptly changes the onsite potential. Previously, spin-1 condensates were quenched to demonstrate intriguing nonequilibrium dynamics \cite{leslie2009PhysRevA.79.043631,lamacraft2007PhysRevLett.98.160404,stamperKurn2013RevModPhys.85.1191}. We characterize this uniform polar condensate with a filling factor $n_B$ for condensed bosons and a chemical potential  $\mu=-3t_1 + c_0 n_B$ at which the energy per lattice site is minimized. We note that such a polar state is energetically favored as long as $q'$ is set to a sufficiently large positive value.

Following the usual practice (see, for example, \cite{kain2014PhysRevA.90.063626}), we apply the Bogoliubov perturbation ansatz to the postquench Hamiltonian where the system parameters are fixed at their quenched values.  When expanded up to second order in $\hat{b}_{\vb{j}}$ = $\hat{b}'_{\vb{j}} - (0,\sqrt{n_B},0)$, which are the field operators describing quantum fluctuations on top of the condensate, we find the postquench Hamiltonian to be block diagonal,
\begin{equation}
	\label{eq:total H in position}
	\hat{H} = \hat{H}_{1/2} \oplus \hat{H}_{0},
\end{equation} where
	\begin{equation}
		\label{eq:H_half}
		\begin{split}
		\hat{H}_{1/2} = &  -t_1 \sum_{\left< \vb{ij} \right>} \hat{b}_{\vb{i}}^\dag s_0 \hat{b}_{\vb{j}} + t_2\sum_{\left<\left< \vb{ij} \right> \right>} e^{-i\nu_{\vb{ij}} \phi} \hat{b}_{\vb{i}}^\dag s_z \hat{b}_{\vb{j}} \\ 
		&+ q \sum_{\vb{i}} \hat{b}_{\vb{i}}^\dag s_0 \hat{b}_{\vb{i}} 
		+ m \sum_{\vb{i}} \xi_{\vb{i}}  \hat{b}_{\vb{i}}^\dag s_0 \hat{b}_{\vb{i}} \\
		&+\frac{c_2n_B}{2} \sum_{\vb{i}}(\hat{b}_{\vb{i}} s_x \hat{b}_{\vb{i}} +\hat{b}^\dag_{\vb{i}} s_x \hat{b}^\dag_{\vb{i}})
		\end{split}
	\end{equation}
is the Hamiltonian in the pseudo spin-$1/2$ space $(\ket{\uparrow}\equiv \ket{S=1,M=+1}, \ket{\downarrow} \equiv \ket{S=1,M=-1})$, where $s_{x,y,z}$ and $s_0$ are the Pauli and identity matrices in spin space and 
\begin{equation}
	q =q' + 3t_1 + n_B c_2.
\end{equation}  
As can be seen, without pairing terms, $\hat{H}_{1/2}$ in Eq. (\ref{eq:H_half}) represents the spinful (doubled) Haldane model. Thus, with pairing terms, it describes what we call a BdG extension of spinful Haldane model.
Note that we have used a lowercase $s$ instead of capital $S$ to distinguish between spin-$1/2$ and spin-$1$ systems. In Eq. (\ref{eq:total H in position}), 
$\hat{H}_0$ (which we have not shown) is the Hamiltonian for the $\ket{S=1,M=0}$ component which always has a stable spectrum (for weak $c_0$) and is not affected by the quench. 
 
 \begin{figure}[h]
 	\centering
 	\includegraphics[width=0.25\textwidth]{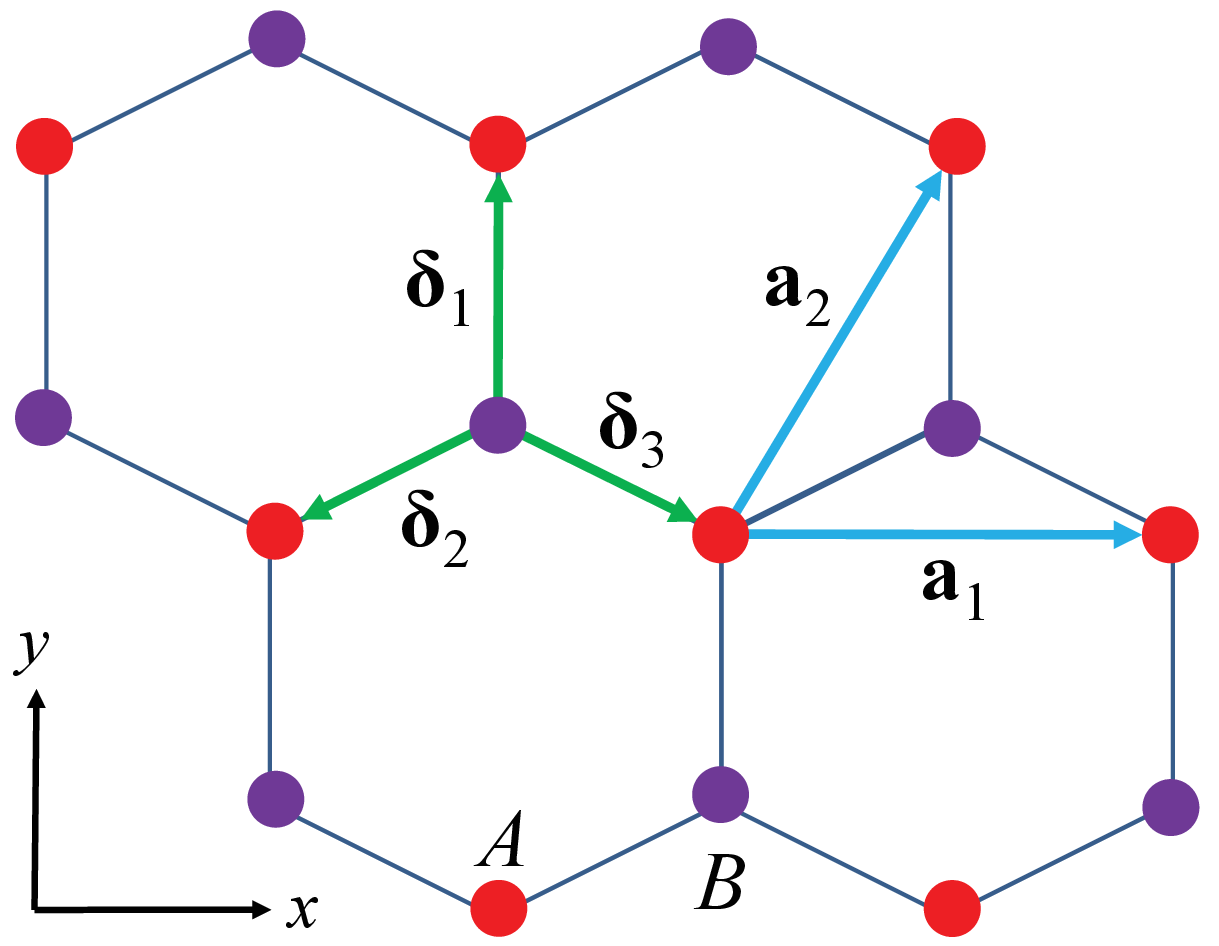}
 	\caption{A honeycomb lattice model where $\vb{a}_1$ and  $\vb{a}_2$ are  basis vectors, and $\delta_1, \delta_2, \delta_3$ are vectors between a site and its three nearest neighbors. The separation between two adjacent sites on the same sublattice is chosen as the distance unit.}
 	\label{fig:a graphene model}
 \end{figure}

We focus exclusively on the spin-$1/2$ system in a honeycomb lattice.  We assume a stripe geometry with open boundaries along $y$ and periodic (zigzag) boundaries along $x$. We apply a partial Fourier transformation along $x$,
\begin{equation}
	\hat{b}_{\vb{j}\equiv(j_x,j_y)} =\sum_{k_x} \hat{b}_{k_x,j_y} \frac{e^{ik_x X_{\vb{j}}}}{\sqrt{N_x}}, j_y = 1, 2, \cdots, N_y,
\end{equation}
where $N_x$ ($N_y$) is the number of unit cells along $x$ ($y$), $k_x$ is the momentum vector along $x$, and $X_{\vb{j}}$ is the $x$-component of the position vector at site $\vb{j}$. Finally, with the help of the basis vectors defined in Fig. \ref{fig:a graphene model}, we change the Hamiltonian (\ref{eq:H_half}) to
\begin{equation}
	\hat{H}_{1/2} =\frac{1}{2}\sum_{k_x}\hat{\psi}^\dag_{k_x} \Sigma_z H_{BdG}(k_x) \hat{\psi}_{k_x},
\end{equation} where $\hat{\psi}_{k_x} = (\hat{b}_{k_x},\hat{b}^\dag_{-k_x})$ is the Nambu spinor and $H_{BdG}(k_x)$ is the bosonic BdG Hamiltonian given by Eq. (\ref{eq:H_BdG}) 
with the matrix $\Sigma_z$ defined directly below
Eq. (\ref{eq:particle-hole symmetry}).  
Here, $\hat{b}_{k_x}$ is the vector field and $A(k_x)$ and $B(k_x)$ in Eq. (\ref{eq:H_BdG}) 
are matrices given by Eq. (\ref{eq:A and B for our model}) 
in the internal space which is now the tensor product between the $y$-unit cell space, the spin space and the sublattice space defined in the main text. For convenience, we change $k_x$ to $k$ and $j_y$ to $j$ in the main text.


%


\begin{thebibliography}{73}
\bibitem [{lin()}]{ling2020PRL.Note1}%
  \BibitemOpen
  \href@noop {} {\bibinfo  {journal} {Although derived from a BdG Hamiltonian
  that depends on crystal momentum $k$, this dependency is not essential.}\
  }\BibitemShut {NoStop}%
\end{thebibliography}

\end{document}